\documentclass[aps,prb,twocolumn,superscriptaddress,showpacs,showkeys]{revtex4}

\usepackage{amsmath}
\usepackage{amssymb}
\usepackage{graphicx}
\usepackage{relsize}
\usepackage{bm}
\usepackage{color}

\renewcommand{\i}{\ensuremath{\mathrm{i}}}
\newcommand{\e}{\ensuremath{\mathrm{e}}}
\renewcommand{\d}{\ensuremath{\mathrm{d}}}

\begin{document}
\title{Effects of optical and surface polar phonons on the optical conductivity of doped graphene}
\author{Benedikt Scharf}
\affiliation{Institute for Theoretical Physics, University of Regensburg, 93040 Regensburg, Germany}
\author{Vasili Perebeinos}
\email[]{vperebe@us.ibm.com}
\affiliation{IBM Research Division T. J. Watson Research Center, Yorktown Heights, New York 10598, USA}
\author{Jaroslav Fabian}
\affiliation{Institute for Theoretical Physics, University of Regensburg, 93040 Regensburg, Germany}
\author{Phaedon Avouris}
\affiliation{IBM Research Division T. J. Watson Research Center, Yorktown Heights, New York 10598, USA}

\date{\today}

\begin{abstract}
Using the Kubo linear response formalism, we study the effects of intrinsic graphene optical and surface polar phonons (SPPs) on the optical conductivity of doped graphene. We find that inelastic electron-phonon scattering contributes significantly to the phonon-assisted absorption in the optical gap. At room temperature, this midgap absorption can be as large as 20-25\% of the universal ac conductivity for graphene on polar substrates (such as Al$_2$O$_3$ or HfO$_2$) due to strong electron-SPP coupling. The midgap absorption, moreover, strongly depends on the substrates and doping levels used. With increasing temperature, the midgap absorption increases, while the Drude peak, on the other hand, becomes broader as inelastic electron-phonon scattering becomes more probable. Consequently, the Drude weight decreases with increasing temperature.
\end{abstract}


\maketitle

\section{Introduction}\label{Sec:Intro}
Since it was first isolated in 2004,\cite{Novoselov2004:Science} graphene, a material composed of a single layer of carbon atoms arranged in a two-dimensional (2D) honeycomb lattice, has attracted immense interest\cite{Novoselov2005:Nature,Geim2007:NatureMat,CastroNeto2009:RMP} due to its excellent transport and optical properties,\cite{Novoselov2005:Nature,Zhang2005:Nature,Bolotin2008:SolidStateComm,Du2008:NatureNano,CastroNeto2009:RMP,Peres2010:RMP,DasSarma2011:RMP} which make it an attractive candidate for possible applications in nanoscale electronics and optoelectronics.\cite{Geim2009:Science,Avouris2010:NL,Ferrari2010:NP}
One particular field which has received considerable attention, both experimentally\cite{Nair2008:Science,Li2008:NaturePhys,Mak2008:PRL,Horng2011:PRB} as well as theoretically,\cite{Peres2006:PRB,Gusynin2006:PRB,Falkovsky2007:PRB,Stauber2008:PRB2,Stauber2008:PRB,Carbotte2010:PRB,Peres2010:PRL,Abedinpour2011:PRB,Vasko2012:arxiv} is the optical (or ac) conductivity in graphene, that is, the frequency-dependent conductivity. The main feature that can be observed in the optical conductivity is that for frequencies larger than twice the absolute value of the chemical potential $\mu$, the optical conductivity is roughly given by $\sigma_0=e^2/(4\hbar)$, the so-called universal ac conductivity.\cite{Nair2008:Science,Li2008:NaturePhys} For frequencies below $2|\mu|$, the optical conductivity is greatly reduced, which can be explained within a single-particle model where transitions induced by photons with energies $\hbar\omega<2|\mu|$ are forbidden due to Pauli's exclusion principle. The ability to tune the optical properties of graphene has been explored for use in broadband light modulators.\cite{Chen2011:NP,Liu2012:NL,Bao2012:ACSNano} One figure of merit for this application is the modulation depth of the optical absorption. In experiments, however, one does not observe the optical conductivity to vanish completely, as one would expect from the simple single-particle argument given above. In addition, a substantial Drude weight loss has been reported in graphene on SiO$_2$.\cite{Horng2011:PRB}

To describe this behavior, mechanisms involving disorder and/or phonons, both of which can account for a finite absorption below $2|\mu|$, have been studied theoretically in both monolayer\cite{Stauber2008:PRB2,Stauber2008:PRB,Carbotte2010:PRB,deJuan2010:PRB,Vasko2012:arxiv} and bilayer\cite{Min2011:PRB,Abergel2012:PRB} graphene. In addition to these single-electron effects, excitonic effects\cite{Peres2010:PRL} as well as effects arising from the Coulomb interaction\cite{Abedinpour2011:PRB} have also been considered, but were found to have a negligible effect on the midgap absorption in heavily doped samples. Moreover, the optical conductivity in the presence of a magnetic field, the so-called magneto-optical conductivity, has also been investigated theoretically,\cite{Gusynin2007:PRL,Pound2012:PRB} with Ref.~\onlinecite{Pound2012:PRB} taking into account the coupling between electrons and Einstein phonons.

Besides the aforementioned studies on the optical conductivity, the role played by different phonons has also been studied in the context of heat dissipation mechanisms\cite{Price2012:PRB,Low2012:PRB,Petrov2012:unpublished} and current/velocity saturation in graphene,\cite{Meric2008:NatureNano} which plays an important role in electronic RF applications\cite{Avouris2010:NL} and also for transport\cite{Perebeinos2009:NL,Chandra2011:PRL} in the similar system of carbon nanotubes. Inelastic scattering either by intrinsic graphene optical phonons\cite{Barreiro2009:PRL} or surface polar phonons\cite{Meric2008:NatureNano,Freitag2009:NanoLetters,Perebeinos2010:PRB,Konar2010:PRB,Li2010:APL,DaSilva2010:PRL} (SPPs) is thought to give rise to the saturation of the current in graphene and affect the low field carrier mobility.\cite{Chen2008:NP,Fratini2008:PRB} However, from transport experiments alone it is difficult to identify the role played by SPPs from the polar substrates because of the complications arising from charge traps which can be populated thermally\cite{Farmer2011:PRB} or by the high electrical fields.\cite{Chiu2010:NanoLetters}

Here, we show that the temperature dependence of the midgap absorption can be significantly stronger in the presence of SPPs as compared to suspended graphene or graphene on a non-polar substrate such as diamond-like carbon. Our main goal in this manuscript is to study the optical conductivity in the presence of phonons. While the impact of optical phonons has been studied in several earlier works,\cite{Stauber2008:PRB2,Stauber2008:PRB,Carbotte2010:PRB} the effect of SPPs on the optical conductivity in graphene has yet to be analyzed. In this paper, we use linear response theory to derive a Kubo formula for the optical conductivity, which is then evaluated for suspended graphene as well as graphene on different polar substrates, where SPPs are present.

%The manuscript is organized as follows: A brief introduction of the model and formalism in Sec.~\ref{Sec:Model} is followed by a presentation of the results obtained for the optical conductivity of several different graphene systems in Sec.~\ref{Sec:Results}. The paper is finally concluded by a brief summary.

\section{Model}\label{Sec:Model}
To describe the electronic (single-particle) band structure of graphene, we use the Dirac-cone approximation, where the Hamiltonian can be written as
\begin{equation}\label{Hamiltonian_electrons_diagonal}
\hat{H}_{\mathrm{e}}=\sum\limits_{\mathbf{k},s,v,\lambda}\lambda\epsilon_{\mathbf{k}}\,\hat{c}^\dagger_{\lambda\mathbf{k}sv}\hat{c}_{\lambda\mathbf{k}sv},
\end{equation}
with $\epsilon_{\mathbf{k}}=\hbar v_\mathrm{\mathsmaller{F}}k$. Here, $\mathbf{k}$, $s$, and $v$ denote the momentum, spin, and valley quantum numbers, $\lambda$ the conduction ($\lambda=+1$) and valence ($\lambda=-1$) bands, $\hat{c}^{\dagger}_{\lambda\mathbf{k}sv}$ and $\hat{c}_{\lambda\mathbf{k}sv}$ the corresponding creation and annihilation operators, and $v_\mathrm{\mathsmaller{F}}\approx 10^8$ cm/s the Fermi velocity in graphene.\cite{CastroNeto2009:RMP}

Since the goal of this work is to study and compare the effects of several different phonons on the optical conductivity of graphene, we need to take into account the interaction with these phonons. A general phononic Hamiltonian reads as
\begin{equation}\label{Hamiltonian_phonons}
\hat{H}_{\mathrm{ph}}=\sum\limits_{\mathbf{q},\Lambda}\hbar\omega_{\mathsmaller{\Lambda}}\left(\mathbf{q}\right)\hat{p}^\dagger_{\mathbf{q}\Lambda}\hat{p}_{\mathbf{q}\Lambda},
\end{equation}
where different phonon branches are labeled as $\Lambda$, the phonon momentum as $\mathbf{q}$, and the corresponding frequencies and creation (annihilation) operators as $\omega_{\mathsmaller{\Lambda}}\left(\mathbf{q}\right)$ and $\hat{p}^\dagger_{\mathbf{q}\Lambda}$ $(\hat{p}_{\mathbf{q}\Lambda})$. Whereas Eqs.~(\ref{Hamiltonian_electrons_diagonal}) and~(\ref{Hamiltonian_phonons}) describe isolated systems of electrons and phonons, respectively, the coupling between those systems is given by
\begin{equation}\label{Hamiltonian_coupling}
\begin{aligned}
\hat{H}_{\mathrm{e-ph}}=\sum\limits_{\lambda\mathbf{k}sv}\sum\limits_{\lambda'\mathbf{q}v'\Lambda}&M_{vv',\Lambda}^{\lambda\lambda'}(\mathbf{k},\mathbf{q})\left(\hat{p}^\dagger_{-\mathbf{q}\Lambda}+\hat{p}_{\mathbf{q}\Lambda}\right)\\
&\times\hat{c}^\dagger_{\lambda(\mathbf{k}+\mathbf{q})sv}\hat{c}_{\lambda'\mathbf{k}sv'},
\end{aligned}
\end{equation}
where $M_{vv',\Lambda}^{\lambda\lambda'}(\mathbf{k},\mathbf{q})$ is the electron-phonon coupling matrix element.\cite{Mahan2000} Hence, the total Hamiltonian of our model reads as
\begin{equation}\label{Hamiltonian}
\hat{H}=\hat{H}_{\mathrm{e}}+\hat{H}_{\mathrm{ph}}+\hat{H}_{\mathrm{e-ph}}.
\end{equation}

In this work, we investigate two different types of phonons that couple to the electrons in graphene: intrinsic graphene optical phonons and SPPs, that is, surface phonons of polar substrates which interact with the electrons in graphene via the electric fields those phonons generate. Intrinsic graphene acoustic phonons are not included in our model because their effect on the optical conductivity is negligible as has been shown in Ref.~\onlinecite{Stauber2008:PRB2}. The dominant electron-optical phonon coupling\cite{Piscanec2004:PRL,Lazzeri2005:PRL,Perebeinos2005:PRL2,Ando2006:JPSJ} is due to longitudinal-optical (LO) and transverse-optical (TO) phonons at the $\Gamma$-point and TO phonon at the $K$-point. In the vicinity of the $\Gamma$ point, the dispersion of both, LO and TO phonons (denoted by $\Gamma_\mathsmaller{\mathrm{LO}}$ and $\Gamma_\mathsmaller{\mathrm{TO}}$), can be approximated by the constant energy $\hbar\omega_\mathsmaller{\Gamma}\approx197$ meV. Near the $K$ and $K'$ points, on the other hand, only the TO phonon (denoted by $K_\mathsmaller{\mathrm{TO}}$) contributes to the electron self-energy and its dispersion in this region can again be assumed as flat, $\hbar\omega_\mathsmaller{K}\approx157$ meV.\footnote{For phonons near the $K$ point, $\mathbf{q}$ in Eqs.~(\ref{Hamiltonian_phonons}), (\ref{Hamiltonian_coupling}), and~(\ref{second_order_self_energy}) should not be understood as the momentum of the phonon, but as the deviation from the momentum given by the $K$ or $K'$ points.} Furthermore, we need to know the products $\tilde{M}_{\Lambda}\equiv\sum\limits_{\tilde{v}}M_{v\tilde{v},\Lambda}^{\lambda\tilde{\lambda}}(\mathbf{k}-\mathbf{q},\mathbf{q})M_{\tilde{v}v',\Lambda}^{\tilde{\lambda}\lambda'}(\mathbf{k},-\mathbf{q})$ of the electron-phonon coupling matrix elements for each phonon branch in order to calculate the electron self-energy. The coupling of both $\Gamma$ phonons is given by\cite{Piscanec2004:PRL,Lazzeri2005:PRL,Ando2006:JPSJ}
\begin{equation}\label{e-ph_matrix_elements_Gamma}
\tilde{M}_{\Gamma_\mathsmaller{\mathrm{LO}}}+\tilde{M}_{\Gamma_\mathsmaller{\mathrm{TO}}}=\frac{\hbar D^2_\mathsmaller{\Gamma}}{2NM_\mathsmaller{\mathrm{c}}\omega_\mathsmaller{\Gamma}}\left(1+\lambda\lambda'\right)\delta_\mathsmaller{vv'},
\end{equation}
where $N$ is the number of unit cells, $M_\mathsmaller{\mathrm{c}}$ the carbon mass, and $D_\mathsmaller{\Gamma}\approx11.2$ eV/{\AA} the strength of the electron-phonon coupling.\cite{Perebeinos2010:PRB} The coupling of the $K_\mathsmaller{\mathrm{TO}}$ phonon mode to the electrons in graphene is twice as large as that of phonons at the $\Gamma$-point\cite{Piscanec2004:PRL,Lazzeri2005:PRL,Perebeinos2005:PRL2} and is described by\footnote{The matrix element $M_{vv',K_\mathsmaller{\mathrm{TO}}}^{\lambda\lambda'}(\mathbf{k},\mathbf{q})\propto\delta_{v,-v'}$ describes intervalley scattering, but does otherwise not depend on the valley quantum number. Thus, the sum $\tilde{M}_{K_\mathsmaller{\mathrm{TO}}}\equiv\sum\limits_{\tilde{v}}M_{v\tilde{v},K_\mathsmaller{\mathrm{TO}}}^{\lambda\tilde{\lambda}}(\mathbf{k}-\mathbf{q},\mathbf{q})M_{\tilde{v}v',K_\mathsmaller{\mathrm{TO}}}^{\tilde{\lambda}\lambda'}(\mathbf{k},-\mathbf{q})\propto\delta_{v,v'}$ is diagonal in the valley quantum number.}
\begin{equation}\label{e-ph_matrix_elements_K}
\tilde{M}_{K_\mathsmaller{\mathrm{TO}}}=\frac{\hbar D^2_\mathsmaller{\Gamma}}{2NM_\mathsmaller{\mathrm{c}}\omega_\mathsmaller{K}}\left[1+\lambda\lambda'-\tilde{\lambda}\left(\lambda\e^{-\i\theta}+\lambda'\e^{\i\theta}\right)\right]\delta_\mathsmaller{vv'},
\end{equation}
where $\theta\equiv\theta_\mathsmaller{\mathbf{k}}-\theta_\mathsmaller{\mathbf{k}-\mathbf{q}}$ and $\theta_\mathsmaller{\mathbf{k}}=\arctan\left(k_x/k_y\right)$. Our choice of the optical phonon deformation potential lies between the LDA results of Refs.~\onlinecite{Piscanec2004:PRL} and~\onlinecite{Lazzeri2005:PRL} and the GW results from Ref.~\onlinecite{Lazzeri2008:PRB}. A smaller electron-intrinsic optical phonon coupling as suggested in Ref.~\onlinecite{Borysenko2010:PRB} would further reduce the absorption below $2|\mu|$ in suspended graphene.

Here, we include SPPs in our model as follows: There are two surface optical (SO) phonons in polar substrates that interact with the electrons in graphene and whose dispersion can again be approximated by substrate-specific, constant frequencies $\omega_\mathsmaller{\mathrm{SO}_1}$ and $\omega_\mathsmaller{\mathrm{SO}_2}$, and their electron-phonon coupling matrix elements read as\cite{Fratini2008:PRB,Konar2010:PRB}
\begin{equation}\label{e-ph_matrix_elements_SPP}
\tilde{M}_{\mathsmaller{\Lambda}}=\frac{\pi^2e^2F_{\mathsmaller{\Lambda}}^2(q)}{NAq}\e^{-2qz_0}\left[1+\lambda\lambda'+\tilde{\lambda}\left(\lambda\e^{-\i\theta}+\lambda'\e^{\i\theta}\right)\right]\delta_\mathsmaller{vv'},
\end{equation}
where $e=|e|$ is the absolute value of the electron charge, $A=3\sqrt{3}a^2/2$ the area of the graphene unit cell, $a\approx1.42$ {\AA} the distance between two carbon atoms, $z_0\approx3.5$ {\AA} the van der Waals distance between the graphene sheet and the substrate, and the Fr\"{o}hlich coupling $F_{\mathsmaller{\Lambda}}^2(q)$ describes the magnitude of the polarization field.\cite{Wang1972:PRB} The Fr\"{o}hlich coupling is given by\cite{Fischetti2001:JoAP,Price2012:PRB}
\begin{equation}\label{Froehlich_SO1}
F_{\mathsmaller{\mathrm{SO}_1}}^2(q)=\frac{\hbar\omega_\mathsmaller{\mathrm{SO}_1}}{2\pi}\left[\frac{1}{\varepsilon_\mathsmaller{\mathrm{i}}+\varepsilon(q)}-\frac{1}{\varepsilon_\mathsmaller{0}+\varepsilon(q)}\right]
\end{equation}
and
\begin{equation}\label{Froehlich_SO2}
F_{\mathsmaller{\mathrm{SO}_2}}^2(q)=\frac{\hbar\omega_\mathsmaller{\mathrm{SO}_2}}{2\pi}\left[\frac{1}{\varepsilon_\mathsmaller{\infty}+\varepsilon(q)}-\frac{1}{\varepsilon_\mathsmaller{\mathrm{i}}+\varepsilon(q)}\right]
\end{equation}
with the optical, intermediate, and static permittivities $\varepsilon_\mathsmaller{\infty}$, $\varepsilon_\mathsmaller{\mathrm{i}}$, and $\varepsilon_\mathsmaller{0}$ of the substrate as well as the static, low temperature dielectric function $\varepsilon(q)=1+2\pi e^2\Pi_{\mathsmaller{\mathrm{g}}}\left(q,\omega=0\right)/(\kappa q)$, where $\kappa$ is the background dielectric constant and $\Pi_{\mathsmaller{\mathrm{g}}}\left(q,\omega\right)$ the polarization function of graphene as calculated in Refs.~\onlinecite{Wunsch2006:NJoP,Hwang2007:PRB}. The dielectric function $\varepsilon(q)$ accounts for the screening of the Coulomb interaction in the graphene sheet above the polar substrate. If the effect of screening is to be disregarded, we set $\varepsilon(q)=1$ in Eqs.~(\ref{Froehlich_SO1}) and~(\ref{Froehlich_SO2}), for which we obtain the bare Fr\"{o}hlich couplings presented in Table~\ref{tab:parameters}.

\begin{table}
\begin{center}
\begin{tabular}{|c||c|c|c|c|c|}
\hline
 & Al$_2$O$_3$$^a$ & h-BN$^b$ & HfO$_2$$^c$ & SiC$^d$ & SiO$_2$$^e$\\
\hline\hline
$\varepsilon_\mathsmaller{0}$ & 12.53 & 5.09 & 22.0 & 9.7 & 3.90\\
\hline
$\varepsilon_\mathsmaller{\mathrm{i}}$ & 7.27 & 4.575 & 6.58 & - & 3.36\\
\hline
$\varepsilon_\mathsmaller{\infty}$ & 3.20 & 4.10 & 5.03 & 6.5 & 2.40\\
\hline
$\hbar\omega_\mathsmaller{\mathrm{SO}_1}$ [meV] & 56.1 & 101.7 & 21.6 & 116.0 & 58.9\\
\hline
$\hbar\omega_\mathsmaller{\mathrm{SO}_2}$ [meV] & 110.1 & 195.7 & 54.2 & - & 156.4\\
\hline
$F^2_{\mathsmaller{\mathrm{SO}_1}}$ [meV] & \; 0.420\;& \; 0.258\;& \; 0.304\;& \; 0.735\;& \; 0.237\;\\
\hline
$F^2_{\mathsmaller{\mathrm{SO}_2}}$ [meV] & \; 2.053\;& \; 0.520\;& \; 0.293\;& \;-\;& \; 1.612\;\\
\hline
\end{tabular}
\end{center}
\caption{Optical, intermediate, and static permittivities as well as frequencies and bare Fr\"{o}hlich couplings for the SPP scattering on the substrates Al$_2$O$_3$, hexagonal BN, HfO$_2$ SiC, and SiO$_2$.\\
$^a$ Refs.~\onlinecite{Fischetti2001:JoAP, Ong2012:arxiv}\\
$^b$ Refs.~\onlinecite{Geick1966:PR, Perebeinos2010:PRB, Ong2012:arxiv}\\
$^c$ Refs.~\onlinecite{Fischetti2001:JoAP,Perebeinos2010:PRB,Ong2012:arxiv}\\
$^d$ Refs.~\onlinecite{Harris1995,Perebeinos2010:PRB}\\
$^e$ Ref.~\onlinecite{Perebeinos2010:PRB}, which uses averages of values from Refs.~\onlinecite{Fischetti2001:JoAP,Fratini2008:PRB,Konar2010:PRB}.}\label{tab:parameters}
\end{table}

Employing standard diagrammatic perturbation theory\cite{Mahan2000,FetterWalecka2003,BruusFlensberg2006} and inserting the specific expressions for the matrix elements $M_{vv',\Lambda}^{\lambda\lambda'}(\mathbf{k},\mathbf{q})$ (for more details, we refer to the Appendix~\ref{Sec:AppendixSE}), we find that, up to first non-vanishing order, the electronic spectral function is diagonal in the four quantum numbers $\lambda$, $\mathbf{k}$, $s$, and $v$ and is given by
\begin{equation}\label{spectral_function}
\begin{aligned}
\mathcal{A}_{\lambda}&\left(k,\omega\right)=\\
&-2\,\mathrm{Im}\left\{\frac{1}{\omega+\i0^\mathsmaller{+}-\left[\lambda\epsilon_{\mathbf{k}}-\mu+\Sigma_\mathsmaller{\lambda}\left(k,\omega+\i0^\mathsmaller{+}\right)\right]/\hbar}\right\},
\end{aligned}
\end{equation}
where $\Sigma_\mathsmaller{\lambda}\left(k,\i\nu_n\right)$ denotes the imaginary-time self-energy at the imaginary (fermionic) frequency $\i\nu_n$. In the lowest order, the contribution to the self-energy due to phonons is just the sum of the contributions from the different phonons $\Lambda$ of our model. Moreover, we include scattering at Coulomb impurities in our model by adding the contribution $\Sigma^{\mathsmaller{\mathrm{Co}}}\left(k\right)=-\i\hbar/[2\tau(k)]$ to the self-energy, where we use the transport scattering time $\tau(k)$ as calculated in Ref.~\onlinecite{Hwang2009:PRB}, and the total self-energy reads as
\begin{equation}\label{self_energy_simple}
\begin{aligned}
\Sigma_{\mathsmaller{\lambda}}\left(k,\i\nu_n\right)=&\Sigma^{\mathsmaller{\mathrm{Co}}}\left(k\right)+\Sigma^{\mathsmaller{\Gamma}}\left(\i\nu_n\right)+\Sigma^{\mathsmaller{K}}\left(\i\nu_n\right)\\
&+\Sigma^{\mathsmaller{\mathrm{SO}_1}}_{\mathsmaller{\lambda}}\left(k,\i\nu_n\right)+\Sigma^{\mathsmaller{\mathrm{SO}_2}}_{\mathsmaller{\lambda}}\left(k,\i\nu_n\right).
\end{aligned}
\end{equation}
The contribution $\Sigma^{\mathsmaller{\mathrm{Co}}}\left(k\right)$ has been added to the self-energy to model the lineshape of the Drude absorption peak. Throughout this work we use the impurity concentration of $n_\mathsmaller{\mathrm{i}}=5\times10^{11}$ cm$^{-2}$, which is low enough to not affect the midgap absorption of graphene on polar substrates significantly.

The imaginary parts of the contributions from the optical phonons at the $\Gamma$ and $K$ points ($\Lambda=\Gamma,K$) to the retarded self-energy depend only on the frequency and have the form
\begin{equation}\label{self_energy_optical}
\begin{aligned}
\mathrm{Im}&\left[\Sigma^{\mathsmaller{\Lambda}}\left(\omega+\i0^\mathsmaller{+}\right)\right]=\\
&\left[n_\mathsmaller{\Lambda}+n_\mathsmaller{\mathrm{FD}}\left(\hbar\omega_\mathsmaller{\Lambda}-\hbar\omega\right)\right]\,g_\mathsmaller{\Lambda}\left(\hbar\omega+\mu-\hbar\omega_\mathsmaller{\Lambda}\right)\\
&+\left[n_\mathsmaller{\Lambda}+n_\mathsmaller{\mathrm{FD}}\left(\hbar\omega_\mathsmaller{\Lambda}+\hbar\omega\right)\right]\,g_\mathsmaller{\Lambda}\left(\hbar\omega+\mu+\hbar\omega_\mathsmaller{\Lambda}\right)
\end{aligned}
\end{equation}
with the functions $g_\mathsmaller{\Lambda}(\epsilon)=-AD^2_\mathsmaller{\Gamma}|\epsilon|/(2M_\mathsmaller{\mathrm{c}}\hbar\omega_\mathsmaller{\Lambda}v^2_\mathrm{\mathsmaller{F}})$.

Similarly, the effect of the two SPP modes ($\Lambda=\mathrm{SO}_1,\mathrm{SO}_2$) is described by
\begin{equation}\label{self_energy_SPP}
\begin{aligned}
\mathrm{Im}&\left[\Sigma^{\mathsmaller{\Lambda}}_{\mathsmaller{\lambda}}\left(k,\omega+\i0^\mathsmaller{+}\right)\right]=\\
&\left[n_\mathsmaller{\Lambda}+n_\mathsmaller{\mathrm{FD}}\left(\hbar\omega_\mathsmaller{\Lambda}-\hbar\omega\right)\right]\,h^{\mathsmaller{\Lambda}}_{\mathsmaller{\lambda}}\left(k,\hbar\omega+\mu-\hbar\omega_\mathsmaller{\Lambda}\right)\\
&+\left[n_\mathsmaller{\Lambda}+n_\mathsmaller{\mathrm{FD}}\left(\hbar\omega_\mathsmaller{\Lambda}+\hbar\omega\right)\right]\,h^{\mathsmaller{\Lambda}}_{\mathsmaller{\lambda}}\left(k,\hbar\omega+\mu+\hbar\omega_\mathsmaller{\Lambda}\right),
\end{aligned}
\end{equation}
where
\begin{equation}\label{function_SPP}
\begin{aligned}
h^{\mathsmaller{\Lambda}}_{\mathsmaller{\lambda}}\left(k,\epsilon\right)=-\frac{\pi e^2}{2(\hbar v_\mathrm{\mathsmaller{F}})^2}\int\limits_{0}^{2\pi}\d\theta\frac{F_{\mathsmaller{\Lambda}}^2(q)\,\e^{-2qz_0}}{q}\left(|\epsilon|+\lambda\epsilon\cos\theta\right)
\end{aligned}
\end{equation}
with $q\equiv\sqrt{\epsilon^2+\epsilon^2_{\mathbf{k}}-2\epsilon\epsilon_{\mathbf{k}}\cos\theta}/(\hbar v_\mathrm{\mathsmaller{F}})$. In Eqs.~(\ref{self_energy_optical}) and~(\ref{self_energy_SPP}), we have introduced the Fermi-Dirac and Bose-Einstein distribution functions, $n_\mathsmaller{\mathrm{FD/BE}}(\epsilon)=1/[\exp(\beta\epsilon)\pm1]$, where $\beta=1/(k_{\mathsmaller{\mathrm{B}}}T)$ (with $T$ and $k_{\mathsmaller{\mathrm{B}}}$ being the temperature and the Boltzmann constant, respectively), $n_\mathsmaller{\Lambda}=n_\mathsmaller{\mathrm{BE}}\left(\hbar\omega_{\mathsmaller{\Lambda}}\right)$, and the chemical potential $\mu=\mu(T)$. In the following, we ignore the effect of polaronic shifts, that is, the real parts of the self-energies and set $\Sigma_{\mathsmaller{\lambda}}\left(k,\omega+\i0^\mathsmaller{+}\right)\equiv\i\,\mathrm{Im}\left[\Sigma_{\mathsmaller{\lambda}}\left(k,\omega+\i0^\mathsmaller{+}\right)\right]$. The total (retarded) self-energy in our model is thus completely imaginary.

As detailed in the Appendix~\ref{Sec:Appendixoptical_conductivity}, the real part of the optical conductivity can be calculated from the spectral function~(\ref{spectral_function}) via the formula\footnote{Equivalent formulas can also be found in Refs.~\onlinecite{Gusynin2006:PRB} and~\onlinecite{Carbotte2010:PRB}.}
\begin{equation}\label{real_conductivity_final_main}
\begin{aligned}
\sigma\left(\omega\right)=&\frac{\sigma_0v^2_\mathrm{\mathsmaller{F}}}{\pi^2\omega}\sum\limits_{\lambda\lambda'}\int\limits_{-\infty}^{\infty}\d\omega'\int\limits_{0}^{\infty}\d k\,k\,\mathcal{A}_{\lambda}\left(k,\omega'\right)\\
&\times\mathcal{A}_{\lambda'}\left(k,\omega'+\omega\right)\left[n_\mathsmaller{\mathrm{FD}}\left(\hbar\omega'\right)-n_\mathsmaller{\mathrm{FD}}\left(\hbar\omega+\hbar\omega'\right)\right],
\end{aligned}
\end{equation}
which includes the universal ac conductivity $\sigma_0=e^2/(4\hbar)$. In the following, we will use Eqs.~(\ref{spectral_function})-(\ref{function_SPP}) to numerically calculate the spectral function, which is in turn used to calculate the real part of the optical conductivity numerically via Eq.~(\ref{real_conductivity_final_main}). Those calculations are conducted for several different substrates (as well as suspended graphene) with the corresponding parameters summarized in Table~\ref{tab:parameters}.

\section{Results}\label{Sec:Results}

\begin{figure}[t]
\centering
\includegraphics*[width=8cm]{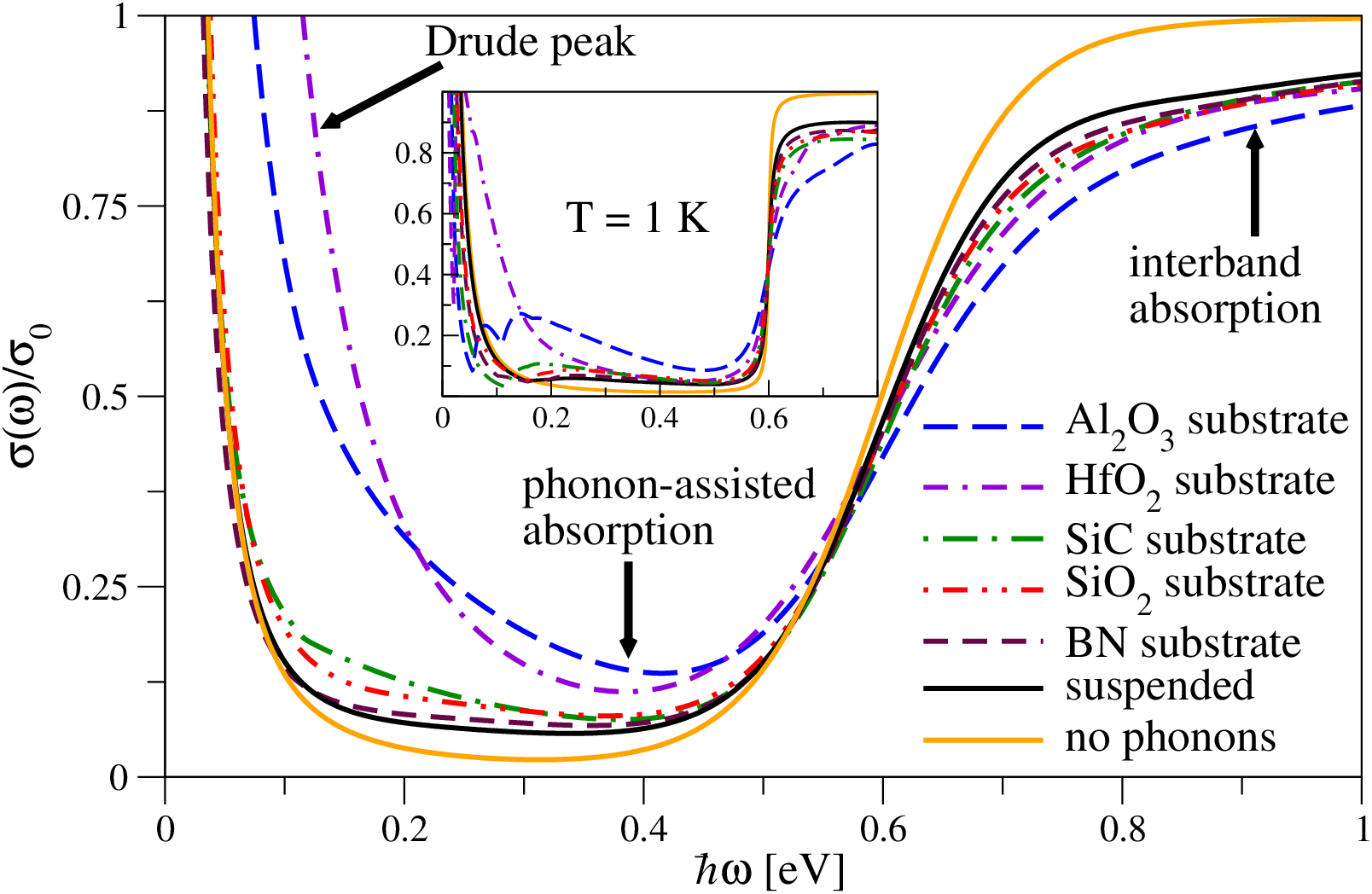}
\caption{(Color online) Calculated frequency dependence of the (real part of the) optical conductivity of suspended graphene and graphene on several different substrates for $n_\mathsmaller{\mathrm{i}}=5\times10^{11}$ cm$^{-2}$ and $\mu=0.3$ eV at $T=300$ K and $T=1$ K (inset).}\label{fig:Conductivity_cp03_different_substrates}
\end{figure}

In Fig.~\ref{fig:Conductivity_cp03_different_substrates}, the optical conductivities\footnote{Our numerical integrations over $\omega$ and $k$ have been conducted on grids with $\Delta(\hbar\omega)=0.25$ meV and $\Delta(\hbar v_\mathsmaller{\mathrm{F}}k)=0.25$ meV. Moreover, to account for the fact that only the imaginary part of the self-energy has been included, the spectral function has been renormalized numerically to $\int\d\omega\mathcal{A}_{\lambda}\left(k,\omega\right)=2\pi$.} (for a fixed chemical potential $\mu=0.3$ eV) at two temperatures $T=1$ K (inset) and $T=300$ K are shown for suspended graphene as well as graphene on several different substrates: Al$_2$O$_3$, hexagonal BN, HfO$_2$, SiC, and SiO$_2$. For comparison, we have also included the optical conductivity of suspended graphene and graphene without any phonon contribution (at $\kappa=1$). Figure~\ref{fig:Conductivity_cp03_different_substrates} has been calculated using the parameters given from Table~\ref{tab:parameters}, the dielectric function $\varepsilon(q)$, and $\kappa=(1+\varepsilon_\mathsmaller{0})/2$ as the background\cite{Hwang2007:PRB,Hwang2009:PRB} dielectric constant.

The profiles in Fig.~\ref{fig:Conductivity_cp03_different_substrates} illustrate the main features that the effect electron-phonon coupling has on the optical conductivity: Whereas there is a gap with a width $2|\mu|$ in the absorption spectrum of the purely electronic single-particle model, where direct transitions between the electronic states in the conduction and valence bands are forbidden for energies $0<\hbar\omega<2|\mu|$ due to Pauli blocking, there is a finite absorption in this region  in the presence of phonons. This finite absorption is largely due to phonon-assisted transitions which give rise to distinct sidebands, the onsets of which can clearly be distinguished from the Drude peak at low temperatures and low impurity densities (see the inset in Fig.~\ref{fig:Conductivity_cp03_different_substrates}). If the photon energy exceeds $2|\mu|$, direct (interband) transitions become possible, resulting in a steep rise of the optical conductivity.

At higher temperatures, one can see that the Drude peak is broadened as more phonons become available and electron-phonon scattering becomes more probable. For HfO$_2$ and Al$_2$O$_3$ substrates, the phonon sidebands merge with the Drude peak resulting in a very broad Drude peak at room temperature. Furthermore, the profiles of the optical conductivity are much smoother compared to those at $T=1$ K and distinct onsets of phonon sidebands can no longer be observed as the profiles of the optical conductivity are smeared out by thermal broadening. Finally, Fig.~\ref{fig:Conductivity_cp03_different_substrates} shows that the so-called ``midgap absorption'', that is, the absorption at $\hbar\omega=\mu$, is significantly enhanced for graphene on polar substrates as compared to suspended graphene or graphene on nonpolar substrates: Whereas the midgap absorption at room temperature is about 5-6\% of $\sigma_0$ for suspended graphene, it can be as high as 20-25\% of $\sigma_0$ for graphene on HfO$_2$ or Al$_2$O$_3$. Hence, the midgap absorption strongly depends on the particular polar substrate used and is, in particular, determined by the interplay between the SPP frequencies $\omega_\mathsmaller{\mathrm{SO}_i}$ and the Fr\"{o}hlich couplings $F^2_{\mathsmaller{\mathrm{SO}_i}}$: the smaller the $\omega_\mathsmaller{\mathrm{SO}_i}$ or the larger $F^2_{\mathsmaller{\mathrm{SO}_i}}$, the larger is the midgap absorption.

Impurity scattering has also an influence on the midgap absorption: By calculating the absorption spectra using just the Coulomb impurity scattering with $\kappa=1$ and no phonons for $\mu=0.3$ eV the Kubo model shows that while at $T=300$ K the midgap absorption is 2.3\% for $n_\mathsmaller{\mathrm{i}}=5\times10^{11}$ cm$^{-2}$ (see also Fig.~\ref{fig:Conductivity_cp03_different_substrates}), it increases to 4.3\% for $n_\mathsmaller{\mathrm{i}}=10^{12}$ cm$^{-2}$. At $T=500$ K, these numbers are 5\% and 7\%, respectively. We note that an obvious first estimate of the midgap absorption could have been obtained by using a Drude model $\sigma(\omega)/\sigma_0=4|\mu|/[\pi\hbar\tau(\omega^2+1/\tau^2)]$, where $\tau$ is a scattering time. For $n_\mathsmaller{\mathrm{i}}=5\times10^{11}$ cm$^{-2}$ and $n_\mathsmaller{\mathrm{i}}=10^{12}$ cm$^{-2}$, the Coulomb scattering mobilities are $\gamma\approx13000$ cm$^2$/(Vs) and $\gamma\approx6500$ cm$^2$/(Vs), respectively, corresponding to midgap absorptions from the Drude model of $\sigma(\mu/\hbar)/\sigma_0\approx4\hbar e v_\mathrm{F}^2/(\pi \gamma\mu^2)=0.7$\% and $1.4$\%, respectively. Thus, this simple estimate using the Drude model significantly underestimates the results obtained from the full calculations. Indeed, the deviations between the estimate from the Drude model and the full Kubo formalism calculation become even more pronounced if phonons are taken into account.

\begin{figure}[t]
\centering
\includegraphics*[width=8cm]{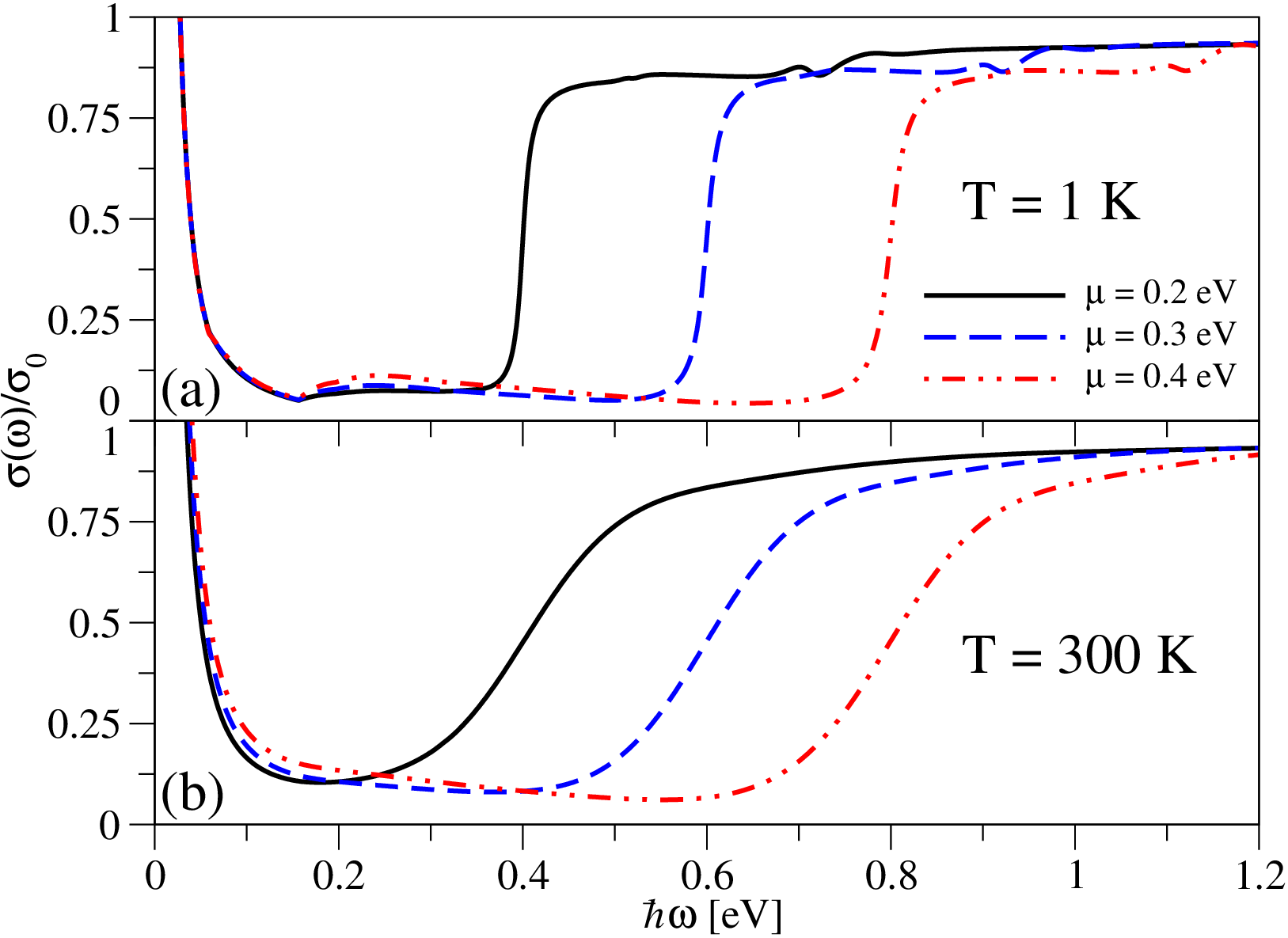}
\caption{(Color online) Calculated frequency dependence of the (real part of the) optical conductivity of graphene on a SiO$_2$ substrate for $n_\mathsmaller{\mathrm{i}}=5\times10^{11}$ cm$^{-2}$ and several different chemical potentials at (a) $T=1$ K and (b) $T=300$ K.}\label{fig:Conductivity_SiO2_different_cps}
\end{figure}

\begin{figure}[t]
\centering
\includegraphics*[width=8cm]{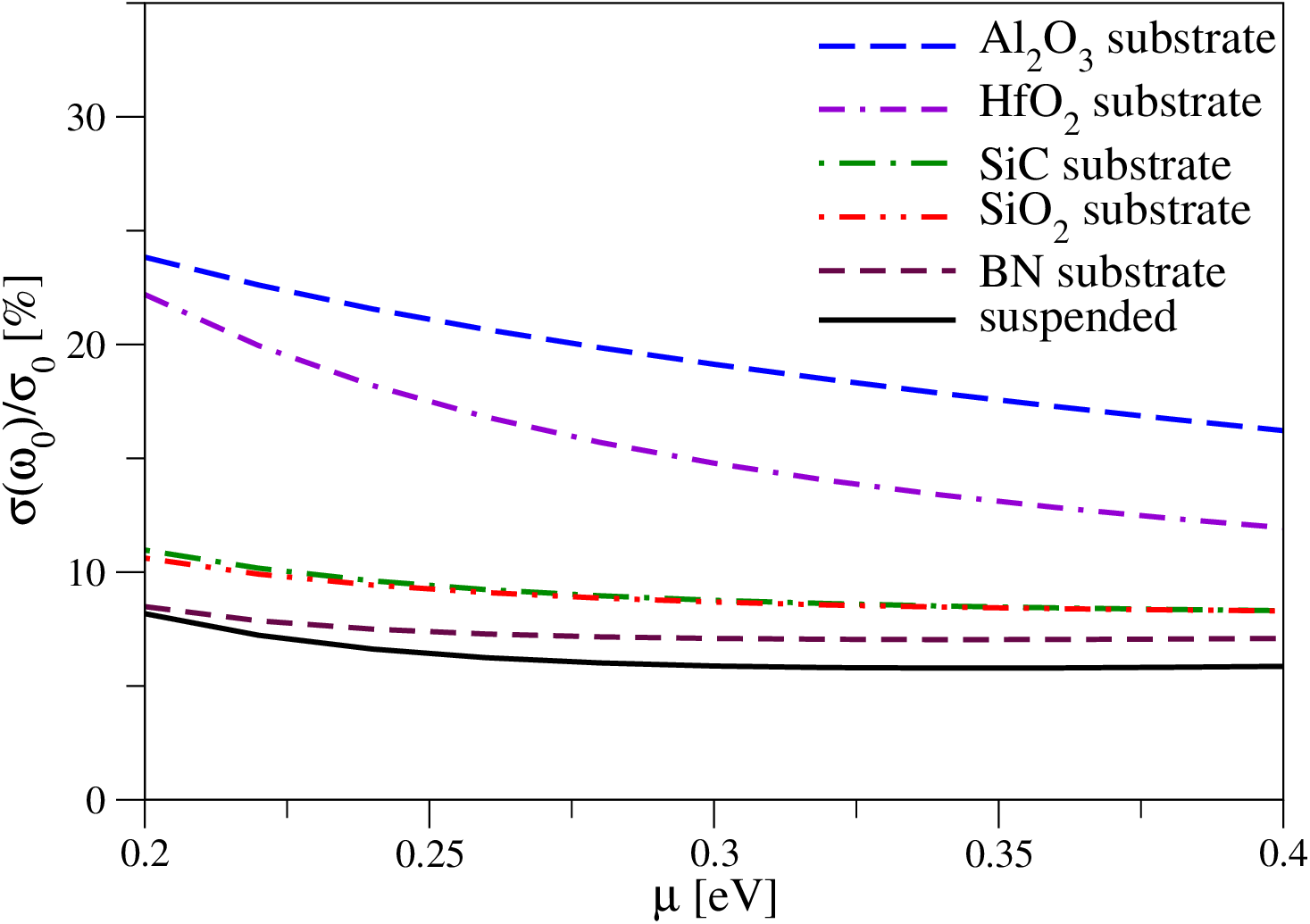}
\caption{(Color online) Calculated dependence of the midgap absorption (at $\hbar\omega_0=\mu$) on the chemical potential for suspended graphene and graphene on several different substrates, $n_\mathsmaller{\mathrm{i}}=5\times10^{11}$ cm$^{-2}$, and $T=300$ K.}\label{fig:Conductivity_SiO2_cp_dependence}
\end{figure}

Figure~\ref{fig:Conductivity_SiO2_different_cps} shows the optical conductivity for graphene on a SiO$_2$ substrate at different temperatures and chemical potentials. Apart from the trends in the behavior of the optical conductivity discussed above, one can clearly see different gaps in the absorption spectrum, given by $2|\mu|$ for each chemical potential. Another feature that can be discerned from Fig.~\ref{fig:Conductivity_SiO2_different_cps} is that the maximal value of the phonon-mediated absorption in the gap increases with increasing chemical potential (doping level). Moreover, we note that due to the electron-hole symmetry of the Dirac Hamiltonian, the profiles of the optical conductivity would look the same for $p$-doped graphene. The dependence of the midgap absorption on the chemical potential at room temperature is shown in Fig.~\ref{fig:Conductivity_SiO2_cp_dependence}, again for suspended graphene and graphene on several different substrates. In the region studied here between $\mu=0.2$ eV and $\mu=0.4$ eV, the midgap absorption decreases with increasing chemical potential for graphene on substrates, with the decay being most pronounced for HfO$_2$. The differences in the shape of the phonon mediated gap absorption reflect the momentum dependence of the electron-phonon interaction in Eq.~(\ref{e-ph_matrix_elements_Gamma})-(\ref{e-ph_matrix_elements_SPP}).

\begin{figure}[t]
\centering
\includegraphics*[width=8cm]{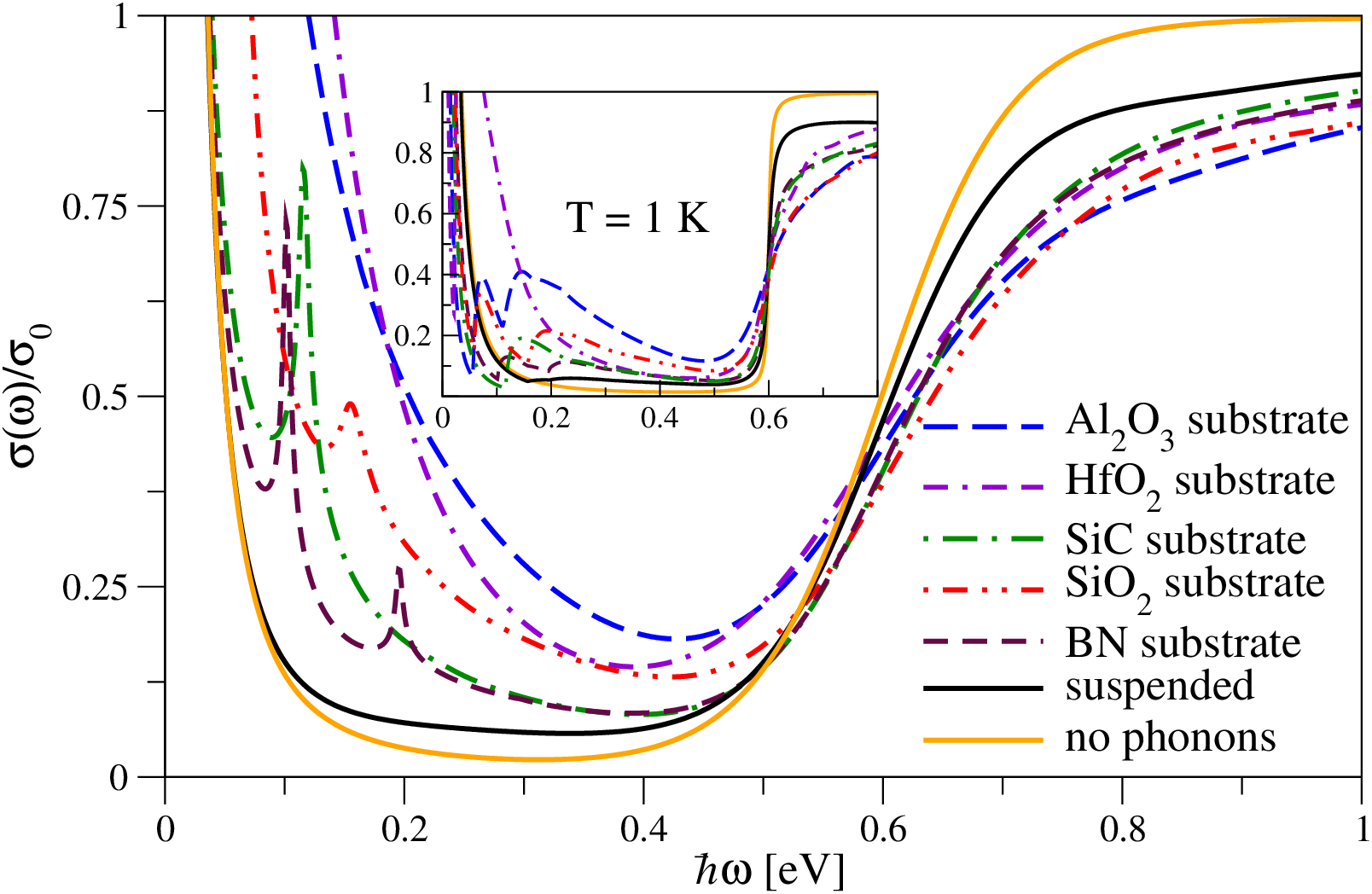}
\caption{(Color online) Calculated frequency dependence of the (real part of the) optical conductivity of suspended graphene and graphene on several different substrates for $n_\mathsmaller{\mathrm{i}}=5\times10^{11}$ cm$^{-2}$ and $\mu=0.3$ eV at $T=300$ K and $T=1$ K (inset) if bare Fr\"{o}hlich couplings are used.}\label{fig:Conductivity_cp03_different_substrates_noscr}
\end{figure}

In particular, Fig.~\ref{fig:Conductivity_cp03_different_substrates_noscr} reveals a striking difference in the absorption if we use a bare unscreened Fr\"{o}hlich couplings. If screening is not accounted for, we find that the optical conductivity in the optical gap is greatly enhanced compared to the situation where screening is used. The most noticeable feature if the bare Fr\"{o}hlich coupling is used, is that a second clearly distinguishable phonon sideband peak (due to the SPPs) can now be observed in the absorption spectra even at room temperature for SiO$_2$ and SiC substrates. For BN substrates, one can even find two such peaks at room temperature. We suggest, therefore, that measurements of the midgap absorption in graphene on different substrates could help to clarify the still controversial\cite{Konar2010:PRB,Li2010:APL,Price2012:PRB,Fischetti2001:JoAP,Ong2012:arxiv} issue concerning the effect of screening on the electron-SPP coupling strength.

\begin{figure}[t]
\centering
\includegraphics*[width=8cm]{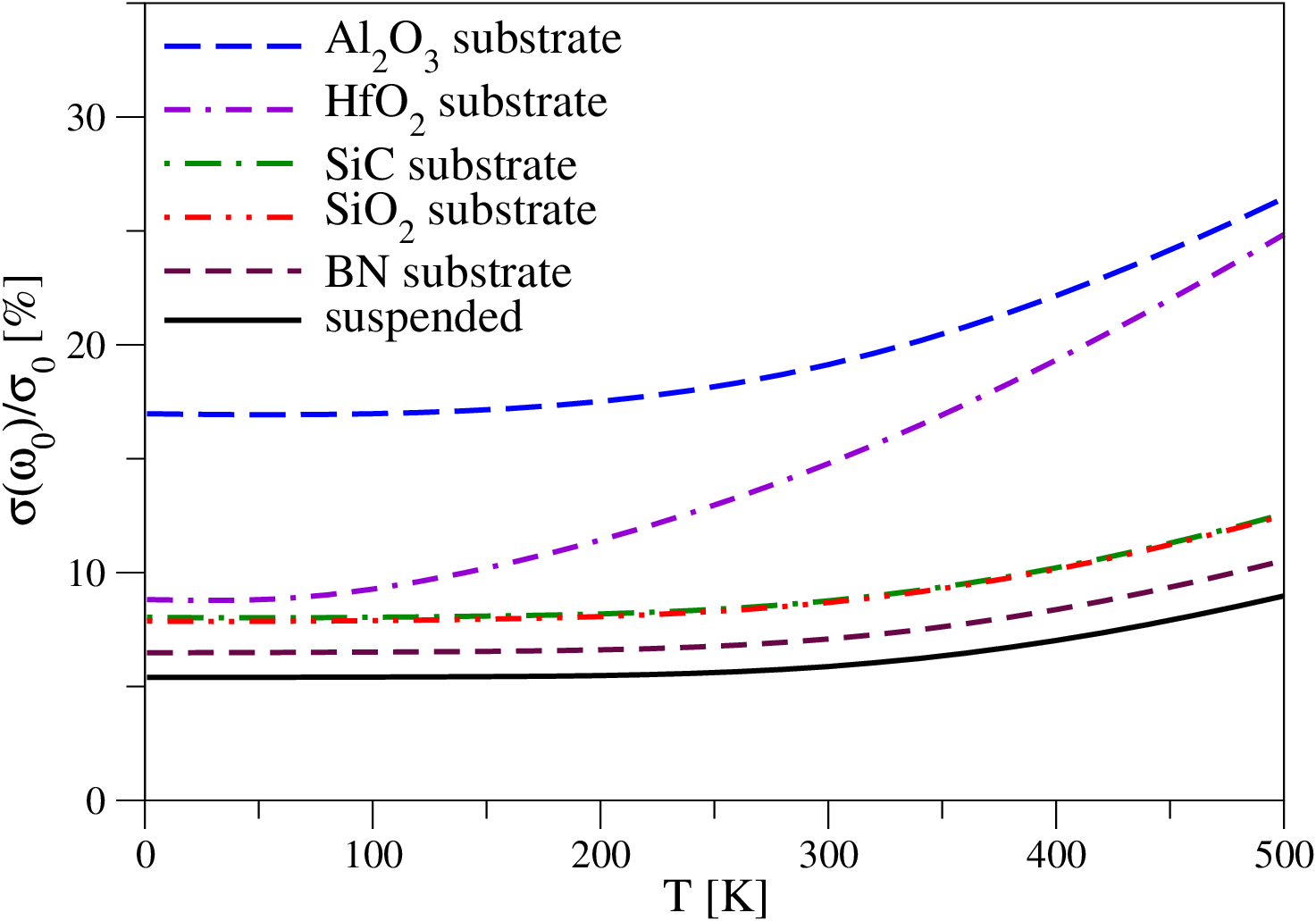}
\caption{(Color online) Calculated temperature dependence of the midgap absorption (at $\hbar\omega_0=\mu$) for suspended graphene and graphene on several different substrates, $n_\mathsmaller{\mathrm{i}}=5\times10^{11}$ cm$^{-2}$, and $\mu=0.3$ eV.}\label{fig:Conductivity_midgap_temp}
\end{figure}

We also investigate the temperature dependence of the midgap conductivity (that is, at $\hbar\omega_0=\mu$) within our model in Fig.~\ref{fig:Conductivity_midgap_temp} on several different substrates. At temperatures below 100 K, the midgap absorption does not depend strongly on the temperature. At about 100 K, an increase of the optical conductivity at $\hbar\omega_0=\mu$ is predicted to take place. Also, the smaller the energy of the dominant phonon contributing to the gap absorption, the stronger is the temperature dependence in Fig.~\ref{fig:Conductivity_midgap_temp}.

Finally, we relate the midgap absorption to the spectral weight of the Drude peak. Describing the graphene optical conductivity in the non-interacting single-particle picture, the spectral weight of the bare Drude peak is $I_0=\int\limits_{0}^{\omega^\mathsmaller{\mathrm{\prime}}}\d\omega\sigma(\omega)=2|\mu|\sigma_0/\hbar=D_0/2$, where $D_0$ is the bare Drude weight and $\omega^\mathsmaller{\mathrm{\prime}}$ is some characteristic frequency much larger than the scattering rate, but smaller than both the lowest energy of the optical phonon $\hbar\omega_\mathsmaller{\mathrm{opt}}$ and $2|\mu|$. In the presence of phonons, the total spectral weight has to be conserved. The spectral weight contribution due to the midgap absorption can be approximated at low temperatures as $I_\mathsmaller{\mathrm{gap}}=\int\limits_{\omega_\mathsmaller{\mathrm{opt}}}^{2|\mu|}\d\omega\sigma(\omega)\approx\alpha\sigma_0(2|\mu|-\hbar\omega_\mathsmaller{\mathrm{opt}})/\hbar$, where $\alpha$ is the averaged value of $\sigma(\omega)$, that is, of the real part of the optical conductivity, in units of $\sigma_0$. If we further assume that the entire spectral weight lost at the Drude peak is transferred to the optical gap and that $\hbar\omega_\mathsmaller{\mathrm{opt}}\ll 2|\mu|$, within this picture the remaining Drude weight $D$ can then be estimated as $D/D_0=(I_0-I_\mathsmaller{\mathrm{gap}})/I_0\approx1-\alpha$. Thus, from this consideration we expect that, as $\alpha$ increases with increasing temperature or strength of the electron-phonon coupling, the Drude weight is reduced.

\begin{figure}[t]
\centering
\includegraphics*[width=8cm]{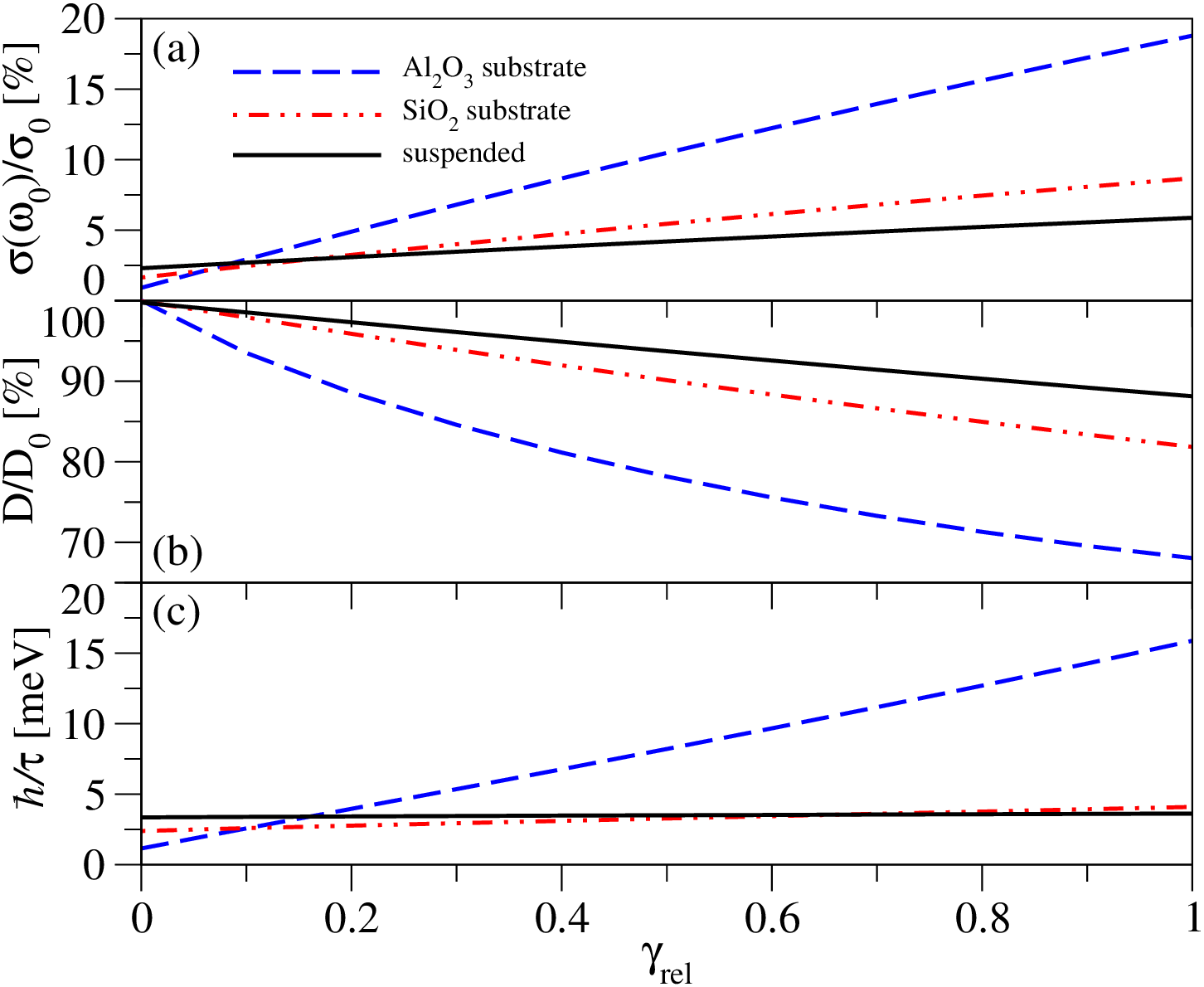}
\caption{(Color online) Calculated dependence of the (a) absorption at $\hbar\omega_0=\mu$, (b) the Drude weight $D$, and (c) the inverse scattering time $\hbar/\tau$ on the relative strength of the electron-phonon coupling $\gamma_\mathsmaller{\mathrm{rel}}$ for suspended graphene and graphene on Al$_2$O$_3$ and SiO$_2$, $n_\mathsmaller{\mathrm{i}}=5\times10^{11}$ cm$^{-2}$, $T=300$ K, and $\mu=0.3$ eV.}\label{fig:el_ph_coupling_screened}
\end{figure}

Figure~\ref{fig:el_ph_coupling_screened} shows (a) the absorption in the optical gap as well as the fitted (b) Drude weight and (c) inverse scattering time for suspended graphene and graphene on Al$_2$O$_3$ and SiO$_2$ substrates with $\mu=0.3$ eV, $T=300$ K, and $n_\mathsmaller{\mathrm{i}}=5\times10^{11}$ cm$^{-2}$ as functions of the relative strength $\gamma_\mathsmaller{\mathrm{rel}}$ of the electron-phonon coupling. Here, the optical conductivities have been calculated by scaling the (products of) electron-phonon coupling matrix elements with $\tilde{M}_{\Lambda}\to\gamma_\mathsmaller{\mathrm{rel}}\tilde{M}_{\Lambda}$, and the Drude weight $D$ as well as the inverse scattering time $1/\tau$ have been extracted from the optical conductivity by fitting the Drude peak to a Lorentzian. As expected from the argument given above, the Drude weight decreases with increasing $\gamma_\mathsmaller{\mathrm{rel}}$, although the simple  relationship between reduced Drude weight loss and the midgap absorption as $1-\alpha$ does not hold. This is because the midgap absorption does not coincide with the averaged value of $\sigma(\omega)$ in the gap and because some of the spectral weight from the Drude peak is transferred not only to the optical gap, but also to the spectral region  $\hbar\omega>2|\mu|$.

With increasing $\gamma_\mathsmaller{\mathrm{rel}}$, electron-phonon scattering becomes more probable and consequently the scattering time decreases as can be seen in Fig.~\ref{fig:el_ph_coupling_screened}~(c). The corresponding increase of $1/\tau$ is most pronounced for graphene on Al$_2$O$_3$ and least pronounced for suspended graphene. Finally, we remind the reader that for $n_\mathsmaller{\mathrm{i}}=5\times10^{11}$ cm$^{-2}$ impurity scattering also contributes to the absorption in the optical gap, which can be discerned from the finite absorption at $\gamma_\mathsmaller{\mathrm{rel}}=0$. Because suspended graphene and graphene on different substrates each possess different background dielectric constants and thus different transport scattering times, the residual values at $\gamma_\mathsmaller{\mathrm{rel}}=0$ are also different.

\section{Conclusions}\label{Sec:Conclusions}
We have studied the effects of intrinsic graphene optical phonons and SPPs on the optical conductivity of doped graphene. Our focus has been on the absorption at frequencies $\hbar\omega<2|\mu|$, where optical transitions are forbidden due to Pauli blocking in a clean system (at $T=0$), but which can occur if phonons are present, giving rise to phonon sidebands. Here, we have found that inelastic phonon scattering contributes significantly to the absorption in the optical gap and strongly depends on the substrate used: At room temperature (and $\mu=0.3$ eV), the midgap absorption, which is mainly due to intrinsic optical phonons, amounts to about 5-6\% of the universal ac conductivity for suspended graphene or graphene on non-polar substrates, while the midgap absorption can be as large as 20-25\% of $\sigma_0$ for graphene on polar substrates (such as Al$_2$O$_3$ or HfO$_2$) due to the smaller SPP energy and strong electron-SPP coupling. Moreover, the midgap absorption depends on the doping level and decreases with increasing $|\mu|$, while the maximal value of the sideband absorption at low temperatures increases.  We have also investigated the temperature dependence of the midgap absorption which increases with increasing temperature. The Drude peak, on the other hand, becomes broader with increasing temperature as inelastic electron-phonon scattering becomes more important. Consequently, the Drude weight decreases with increasing temperature due to the stronger phonon coupling.

\acknowledgments
We gratefully acknowledge Fengnian Xia from IBM for stimulating discussions and Alex Matos-Abiague and Christian Ertler from the University of Regensburg for technical discussions. B.S. and J.F. acknowledge support from DFG Grants No. SFB 689 and No. GRK 1570.

\appendix

\section{Self-energy and Green's Function}\label{Sec:AppendixSE}
We use standard diagrammatic perturbation theory (with the unperturbed Hamiltonian $\hat{H}_{\mathrm{e}}+\hat{H}_{\mathrm{ph}}$ and the perturbation $\hat{H}_{\mathrm{e-ph}}$) to calculate the electronic Matsubara Green's function of the system described by Eq.~(\ref{Hamiltonian}),
\begin{equation}\label{Greens_Function}
\mathcal{G}^{\mathsmaller{\lambda\lambda'}}_{\mathsmaller{vv'}}\left(\mathbf{k},\i\nu_n\right)=-\int\limits_{0}^{\hbar\beta}\d\tau\left\langle\mathcal{T}\left[\hat{c}_{\lambda\mathbf{k}sv}(\tau)\hat{c}^\dagger_{\lambda'\mathbf{k}sv}(0)\right]\right\rangle\e^{\i\nu_n\tau},
\end{equation}
where $\tau$ and $\i\nu_n$ denote the imaginary time and (fermionic) frequency, $\langle...\rangle$ the thermal average, $\mathcal{T}$ the imaginary time-ordering operator, and $\beta=1/(k_{\mathsmaller{\mathrm{B}}}T)$.\cite{Mahan2000,FetterWalecka2003,BruusFlensberg2006} By solving the corresponding Dyson equation, we can express the electronic Green's function via the self-energy $\Sigma_\mathsmaller{\lambda\lambda',vv'}\left(\mathbf{k},\i\nu_n\right)$. In writing down Eq.~(\ref{Greens_Function}), we have used that, due to the conservation of momentum and spin, the self-energy and thus the Green's function are diagonal in $\mathbf{k}$ and $s$ and do not depend on $s$ due to spin degeneracy.

Up to the first non-vanishing order (and omitting the tadpole diagram, which yields a purely real self-energy contribution that can be absorbed in the chemical potential), we find the electronic self-energy due to phonons for arbitrary matrix elements $M_{vv',\Lambda}^{\lambda\lambda'}(\mathbf{k},\mathbf{q})$ to be\cite{Mahan2000}
\begin{equation}\label{second_order_self_energy}
\begin{aligned}
\Sigma^{\mathsmaller{\mathrm{ph}}}_{\mathsmaller{\lambda\lambda',vv'}}&\left(\mathbf{k},\i\nu_n\right)\approx\Sigma^{(2)}_\mathsmaller{\lambda\lambda',vv'}\left(\mathbf{k},\i\nu_n\right)\\
&=\frac{1}{\hbar}\sum\limits_{\Lambda,\mathbf{q},\tilde{\lambda},\tilde{v}}M_{v\tilde{v},\Lambda}^{\lambda\tilde{\lambda}}(\mathbf{k}-\mathbf{q},\mathbf{q})M_{\tilde{v}v',\Lambda}^{\tilde{\lambda}\lambda'}(\mathbf{k},-\mathbf{q})\\
&\quad\times\Biggl[\frac{n_\mathsmaller{\mathbf{q}\Lambda}+1-f_\mathsmaller{\mathrm{FD}}(\tilde{\lambda}\epsilon_{|\mathbf{k}-\mathbf{q}|})}{\i\nu_n-(\tilde{\lambda}\epsilon_{|\mathbf{k}-\mathbf{q}|}-\mu)/\hbar-\omega_{\mathsmaller{\Lambda}}(\mathbf{q})}\\
&\quad\quad\quad+\frac{n_\mathsmaller{\mathbf{q}\Lambda}+f_\mathsmaller{\mathrm{FD}}(\tilde{\lambda}\epsilon_{|\mathbf{k}-\mathbf{q}|})}{\i\nu_n-(\tilde{\lambda}\epsilon_{|\mathbf{k}-\mathbf{q}|}-\mu)/\hbar+\omega_{\mathsmaller{\Lambda}}(\mathbf{q})}\Biggr]
\end{aligned}
\end{equation}
with the Fermi-Dirac and Bose-Einstein distribution functions, $f_\mathsmaller{\mathrm{FD}}\left(\epsilon\right)=n_\mathsmaller{\mathrm{FD}}\left(\epsilon-\mu\right)$ and $n_\mathsmaller{\mathbf{q}\Lambda}=n_\mathsmaller{\mathrm{BE}}\left[\hbar\omega_{\mathsmaller{\Lambda}}(\mathbf{q})\right]$, and the chemical potential $\mu=\mu(T)$ at the temperature $T$. Thus, in the lowest order, the self-energy is simply the sum of the contributions from the different phonons $\Lambda$.

In order to calculate the self-energy, we need to know the products of the electron-phonon coupling matrix elements entering Eq.~(\ref{second_order_self_energy}), $\sum\limits_{\tilde{v}}M_{v\tilde{v},\Lambda}^{\lambda\tilde{\lambda}}(\mathbf{k}-\mathbf{q},\mathbf{q})M_{\tilde{v}v',\Lambda}^{\tilde{\lambda}\lambda'}(\mathbf{k},-\mathbf{q})$, for the $\Gamma$, $K$, SO$_1$, and SO$_2$ modes. Since $n_\mathsmaller{\mathbf{q}\Gamma_\mathsmaller{\mathrm{TO}}}=n_\mathsmaller{\mathbf{q}\Gamma_\mathsmaller{\mathrm{LO}}}$, it follows from Eqs.~(\ref{second_order_self_energy}) and~(\ref{e-ph_matrix_elements_Gamma}) that the contribution due to the optical phonons near the $\Gamma$ point is diagonal in the valley and band quantum numbers. We calculate the self-energy by using the transformation $\mathbf{k}'=\mathbf{k}-\mathbf{q}$ and replacing the sum $\sum\limits_{\mathbf{k}'}$ by the 2D integral $S/(2\pi)^2\int\d k'k'\int\d\theta$, where $S=NA$ and $\theta$ is chosen to be the angle between $\mathbf{k}$ and $\mathbf{k}'$.

Equation~(\ref{e-ph_matrix_elements_K}) describes the coupling of the $K_\mathsmaller{\mathrm{TO}}$ mode to the electrons in graphene. As above, the self-energy contribution from the $K$ phonons is calculated by introducing $\mathbf{k}'$ and writing the sum as a 2D integral. After performing the integration over the angle $\theta$, the terms containing $\e^{\pm\i\theta}$ in Eq.~(\ref{e-ph_matrix_elements_K}) vanish and consequently the contribution to the self-energy is diagonal with respect to the band quantum number. Moreover, Eqs.~(\ref{e-ph_matrix_elements_K}) and~(\ref{second_order_self_energy}) make it clear that, even though the matrix element $M_{vv',K_\mathsmaller{\mathrm{TO}}}^{\lambda\lambda'}(\mathbf{k},\mathbf{q})$ describes intervalley scattering, the second-order contribution from the $K$ phonon to the self-energy is diagonal also in the valley quantum number.

Using the SPP coupling matrix elements~(\ref{e-ph_matrix_elements_SPP}), the contribution from each SPP to the self-energy is again calculated by introducing $\mathbf{k}'$ and writing the sum as a 2D integral. Then, the angular integration for the offdiagonal elements with respect to the band indices $\lambda$ and $\lambda'$ is of the type $\int\limits_0^{2\pi}\d\theta\;\sin\theta f(\cos\theta)$, where the function $f$ depends only on $\cos\theta$ and on the SPP considered, and vanishes for screened as well as unscreened SPPs. Thus, the lowest order contribution from each SPP to the self-energy is also diagonal in the band and valley quantum numbers.

Combining the results discussed so far, the total contribution from all phonons is given by
\begin{equation}\label{self_energy_phonons}
\Sigma^{\mathsmaller{\mathrm{ph}}}_{\mathsmaller{\lambda\lambda',vv'}}\left(\mathbf{k},\i\nu_n\right)=\delta_\mathsmaller{vv'}\delta_\mathsmaller{\lambda\lambda'}
\Sigma^{\mathsmaller{\mathrm{ph}}}_{\mathsmaller{\lambda}}\left(k,\i\nu_n\right),
\end{equation}
where $\Sigma^{\mathsmaller{\mathrm{ph}}}_{\mathsmaller{\lambda}}\left(k,\i\nu_n\right)$ is given by
\begin{equation}\label{self_energy_phonons_simple}
\begin{aligned}
\Sigma^{\mathsmaller{\mathrm{ph}}}_{\mathsmaller{\lambda}}\left(k,\i\nu_n\right)=&\Sigma^{\mathsmaller{\Gamma}}\left(\i\nu_n\right)+\Sigma^{\mathsmaller{K}}\left(\i\nu_n\right)\\
&+\Sigma^{\mathsmaller{\mathrm{SO}_1}}_{\mathsmaller{\lambda}}\left(k,\i\nu_n\right)+\Sigma^{\mathsmaller{\mathrm{SO}_2}}_{\mathsmaller{\lambda}}\left(k,\i\nu_n\right),
\end{aligned}
\end{equation}
and each individual contribution is calculated from Eq.~(\ref{second_order_self_energy}) as described above for $\lambda=\lambda'$ and $v=v'$. Here, we find that, in contrast to the contributions $\Sigma^{\mathsmaller{\mathrm{SO}_1}}_{\mathsmaller{\lambda}}\left(k,\i\nu_n\right)$ and $\Sigma^{\mathsmaller{\mathrm{SO}_2}}_{\mathsmaller{\lambda}}\left(k,\i\nu_n\right)$, the contributions from the graphene optical phonons, $\Sigma^{\mathsmaller{\Gamma}}\left(\i\nu_n\right)$ and $\Sigma^{\mathsmaller{K}}\left(\i\nu_n\right)$, do not depend on the band or momentum $\mathbf{k}$.

The contribution due to Coulomb impurity scattering reads as
\begin{equation}\label{impurity_self_energy_general}
\Sigma^{\mathsmaller{\mathrm{Co}}}_{\mathsmaller{\lambda\lambda',vv'}}(k)\equiv\delta_\mathsmaller{vv'}\delta_\mathsmaller{\lambda\lambda'}\Sigma^{\mathsmaller{\mathrm{Co}}}(k),
\end{equation}
where
\begin{equation}\label{Coulomb_impurity_scattering_self_energy}
\begin{aligned}
\Sigma^{\mathsmaller{\mathrm{Co}}}\left(k\right)=\frac{-\i\hbar}{2\tau(k)}=-\frac{\i\pi n_\mathsmaller{\mathrm{i}}}{2}\int&\frac{\d^2k'}{(2\pi)^2}\left|\frac{2\pi e^2}{\kappa q\varepsilon(q)}\right|^2\delta(\epsilon_k-\epsilon_{k'})\\
&(1-\cos\theta)(1+\cos\theta)
\end{aligned}
\end{equation}
with $\theta\equiv\theta_\mathsmaller{\mathbf{k}}-\theta_\mathsmaller{\mathbf{k}'}$, $\mathbf{q}=\mathbf{k}-\mathbf{k}'$, the dielectric function $\varepsilon(q)$, and the impurity concentration $n_\mathsmaller{\mathrm{i}}$.\cite{Hwang2009:PRB} Since this contribution is also diagonal, the total self-energy
\begin{equation}\label{total_self_energy_general}
\begin{aligned}
\Sigma_\mathsmaller{\lambda\lambda',vv'}\left(\mathbf{k},\i\nu_n\right)&=\delta_\mathsmaller{vv'}\delta_\mathsmaller{\lambda\lambda'}\left[-\frac{\i\hbar}{2\tau(k)}+\Sigma^{\mathsmaller{\mathrm{ph}}}_{\mathsmaller{\lambda}}\left(k,\i\nu_n\right)\right]\\
&\equiv\delta_\mathsmaller{vv'}\delta_\mathsmaller{\lambda\lambda'}\Sigma_{\mathsmaller{\lambda}}\left(k,\i\nu_n\right)
\end{aligned}
\end{equation}
and, consequently, the Green's function
\begin{equation}\label{Greens_Function_via_self_energy}
\begin{aligned}
\mathcal{G}^{\mathsmaller{\lambda\lambda'}}_{\mathsmaller{vv'}}\left(\mathbf{k},\i\nu_n\right)&=\frac{\delta_\mathsmaller{vv'}\delta_\mathsmaller{\lambda\lambda'}}{\i\nu_n-\left[\lambda\epsilon_{\mathbf{k}}-\mu+\Sigma_\mathsmaller{\lambda}\left(k,\i\nu_n\right)\right]/\hbar}\\
&\equiv\delta_\mathsmaller{vv'}\delta_\mathsmaller{\lambda\lambda'}\mathcal{G}_{\mathsmaller{\lambda}}\left(k,\i\nu_n\right)
\end{aligned}
\end{equation}
are diagonal with respect to $\lambda$ and $v$ in our model. Finally, the spectral function is obtained from the Green's function via
\begin{equation}\label{spectral_function_general}
\mathcal{A}_{\lambda}\left(k,\omega\right)=-2\mathrm{Im}\left[\mathcal{G}_{\mathsmaller{\lambda}}\left(k,\omega+\i0^\mathsmaller{+}\right)\right].
\end{equation}

In this work, we are interested only in the imaginary parts of the retarded self-energy. Upon replacing $\i\nu_n$ by $\omega+\i0^\mathsmaller{+}$ in Eq.~(\ref{second_order_self_energy}), the imaginary part of each contribution $\Lambda$ in Eq.~(\ref{second_order_self_energy}) contains a Dirac-$\delta$ function [since there is no contribution from $\mathrm{Im}(\tilde{M}_{\Lambda})$ as discussed above]. After introducing $\mathbf{k}'$ and writing the sum as a 2D integral, the Dirac-$\delta$ function can be used to calculate the $k'$ integral, which then yields Eq.~(\ref{self_energy_optical}) for the $\Gamma$ and $K$ phonons and Eqs.~(\ref{self_energy_SPP}) and~(\ref{function_SPP}) for the SO$_1$ and SO$_2$ phonons.

\section{Kubo formula for the optical conductivity}\label{Sec:Appendixoptical_conductivity}
\subsection{Current density operator}\label{Sec:Appendixcurrent}
Our starting point in the derivation of a Kubo formula for the optical conductivity is the current operator. In the presence of an arbitrary magnetic vector potential $\mathbf{A}(\mathbf{r})$, the (first-quantized) 2D Dirac Hamiltonian of graphene reads as\cite{CastroNeto2009:RMP}
\begin{equation}\label{general_graphene_Hamiltonian}
\hat{H}_{\mathrm{e}}=v_\mathrm{\mathsmaller{F}}\bm{\gamma}.\hat{\bm{\pi}}
\end{equation}
with $\hat{\bm{\pi}}=\hat{\mathbf{p}}+e\mathbf{A}(\mathbf{r})$ being the 2D kinetic momentum operator, $\hat{\mathbf{p}}$ the 2D momentum operator, and the matrices $\gamma_x=\sigma_x\otimes\mathbf{1}$, $\gamma_y=\sigma_y\otimes\tau_z$, $\gamma_z=0$, where $\mathbf{1}$ is the $2\times2$ unity matrix and $\bm{\sigma}$ and $\bm{\tau}$ are Pauli matrices referring to the $A/B$ sublattices and the $K/K'$ points, respectively. As discussed in Ref.~\onlinecite{CastroNeto2009:RMP}, the 2D momentum $\mathbf{k}$ and the valley $K/K'$ are good quantum numbers and the Hamiltonian~(\ref{general_graphene_Hamiltonian}) has the (valley-degenerate) eigenvalues
\begin{equation}\label{graphene_eigenvalues}
\epsilon_{\lambda}(\mathbf{k})=\lambda\epsilon_{\mathbf{k}}=\lambda\hbar v_\mathrm{\mathsmaller{F}}k
\end{equation}
and the corresponding eigenstates
\begin{equation}\label{states_K}
\Psi^{\mathsmaller{\lambda}}_{\mathsmaller{K,\mathbf{k}}}(\mathbf{r})=\frac{\e^{\i\mathbf{k}.\mathbf{r}}}{\sqrt{2S}}\left(\begin{array}{c}
         \e^{-\i\theta_\mathsmaller{\mathbf{k}}/2}\\
         \lambda\e^{\i\theta_\mathsmaller{\mathbf{k}}/2}\\
         \end{array}\right)\otimes\chi_\mathsmaller{K}
\end{equation}
near the $K$ point and
\begin{equation}\label{states_Kp}
\Psi^{\mathsmaller{\lambda}}_{\mathsmaller{K',\mathbf{k}}}(\mathbf{r})=\frac{\e^{\i\mathbf{k}.\mathbf{r}}}{\sqrt{2S}}\left(\begin{array}{c}
         \e^{\i\theta_\mathsmaller{\mathbf{k}}/2}\\
         \lambda\e^{-\i\theta_\mathsmaller{\mathbf{k}}/2}\\
         \end{array}\right)\otimes\chi_\mathsmaller{K'}
\end{equation}
near the $K'$ point, where $S=NA$ denotes the surface area of the graphene sample, $\lambda=\pm1$, $\theta_\mathsmaller{\mathbf{k}}=\arctan\left(k_x/k_y\right)$, and
\begin{equation}\label{spinorsKKp}
\chi_\mathsmaller{K}=\left(\begin{array}{c}1\\0\\\end{array}\right),\quad\chi_\mathsmaller{K'}=\left(\begin{array}{c}0\\1\\\end{array}\right).
\end{equation}

For an arbitrary (normalized) state $\Psi(\mathbf{r})$, the energy expectation value as a functional of the vector potential $\mathbf{A}(\mathbf{r})$ is given by
\begin{equation}\label{energy_functional}
E\left[\mathbf{A}\right]=\sum\limits_{\alpha\beta}\int\d^2r\;\Psi_\alpha^*(x,y)\left(\hat{H}_{\mathrm{e}}\right)_{\alpha\beta}\Psi_\beta(x,y),
\end{equation}
where the sums over $\alpha$ and $\beta$ refer to the matrix $\bm{\gamma}$. The charge current density $\mathbf{j}(\mathbf{r})$ of this state $\Psi(\mathbf{r})$ can be determined by a variational method:
\begin{equation}\label{energy_variation}
\delta E=E\left[\mathbf{A}+\delta\mathbf{A}\right]-E\left[\mathbf{A}\right]=-\int\d^2r\;\mathbf{j}(\mathbf{r})\delta\mathbf{A}(\mathbf{r}),
\end{equation}
which yields
\begin{equation}\label{current_density_1stquant}
\hat{\mathbf{j}}(\mathbf{r})=-ev_\mathrm{\mathsmaller{F}}\sum\limits_{\alpha\beta}\Psi_\alpha^*(\mathbf{r})\left(\bm{\gamma}\right)_{\alpha\beta}\Psi_\beta(\mathbf{r}).
\end{equation}

Promoting the wave functions in Eq.~(\ref{current_density_1stquant}) to field operators, using the eigenbasis given by Eqs.~(\ref{graphene_eigenvalues})-(\ref{states_Kp}), and taking into account the spin degeneracy, the charge current density operator can be determined as
\begin{equation}\label{current_operator_reciprocal_space}
\hat{\mathbf{j}}(\mathbf{q})=\sum\limits_{\mathbf{k}\lambda\lambda'sv}\mathbf{d}^{\mathsmaller{v}}_{\mathsmaller{\lambda\lambda'}}(\mathbf{k},\mathbf{q})\hat{c}^\dagger_{\lambda\mathbf{k}sv}\hat{c}_{\lambda'(\mathbf{k}+\mathbf{q})sv}
\end{equation}
in reciprocal space and as
\begin{equation}\label{current_operator_real_space}
\hat{\mathbf{j}}(\mathbf{r})=\frac{1}{S}\sum\limits_{\mathbf{q}}\e^{\i\mathbf{q}.\mathbf{r}}\hat{\mathbf{j}}(\mathbf{q}),
\end{equation}
in real space. Here, the dipole matrix elements read as
\begin{equation}\label{dipole_elements}
\begin{aligned}
d^{\mathsmaller{K}}_{\mathsmaller{\lambda\lambda',x}}(\mathbf{k},\mathbf{q})=\frac{-ev_\mathrm{\mathsmaller{F}}}{2}\left[\lambda'\e^{\i(\theta_\mathsmaller{\mathbf{k}}+\theta_{\mathsmaller{\mathbf{k}+\mathbf{q}}})/2}+\lambda\e^{-\i(\theta_\mathsmaller{\mathbf{k}}+\theta_{\mathsmaller{\mathbf{k}+\mathbf{q}}})/2}\right],\\
d^{\mathsmaller{K}}_{\mathsmaller{\lambda\lambda',y}}(\mathbf{k},\mathbf{q})=\frac{\i ev_\mathrm{\mathsmaller{F}}}{2}\left[\lambda'\e^{\i(\theta_\mathsmaller{\mathbf{k}}+\theta_{\mathsmaller{\mathbf{k}+\mathbf{q}}})/2}-\lambda\e^{-\i(\theta_\mathsmaller{\mathbf{k}}+\theta_{\mathsmaller{\mathbf{k}+\mathbf{q}}})/2}\right],\\
d^{\mathsmaller{K'}}_{\mathsmaller{\lambda\lambda',x/y}}(\mathbf{k},\mathbf{q})=\left[d^{\mathsmaller{K}}_{\mathsmaller{\lambda\lambda',x/y}}(\mathbf{k},\mathbf{q})\right]^*.
\end{aligned}
\end{equation}

\subsection{Kubo formula}\label{Sec:AppendixKubo}
If an external electric field
\begin{equation}\label{connection_vectorpotential_electricfield}
\mathbf{E}(\mathbf{r},t)=-\frac{\partial\mathbf{A}(\mathbf{r},t)}{\partial t}
\end{equation}
is applied to the system considered here, its effect can be described by
\begin{equation}\label{external_Hamiltonian}
\hat{H}_{\mathrm{ext}}(t)=-\int\d^2r\;\hat{\mathbf{j}}(\mathbf{r}).\mathbf{A}(\mathbf{r},t)
\end{equation}
with the charge current density operator~(\ref{current_operator_real_space}). The total Hamiltonian of the problem then reads as $\hat{H}+\hat{H}_{\mathrm{ext}}(t)$, where $\hat{H}$ is given by Eq.~(\ref{Hamiltonian}).

Using linear response theory [for the unperturbed Hamiltonian $\hat{H}$ and the perturbation $\hat{H}_{\mathrm{ext}}(t)$] and conducting a Fourier transformation with respect to the time and position,\cite{Mahan2000,FetterWalecka2003,BruusFlensberg2006} we find that the current density due to the external field is given by
\begin{equation}\label{current_density}
\delta\langle\hat{j}_\alpha(\mathbf{q},\omega)\rangle=-\frac{1}{\hbar}\sum\limits_{\beta}\Pi_{\alpha\beta}^R\left(\mathbf{q},\omega\right)A_\beta(\mathbf{q},\omega)
\end{equation}
with $\Pi_{\alpha\beta}^R\left(\mathbf{q},\omega\right)$ being the (Fourier transformed) retarded current-current correlation function and $\alpha$ and $\beta$ referring to the in-plane coordinates $x$ and $y$. The retarded correlation function $\Pi_{\alpha\beta}^R\left(\mathbf{q},\omega\right)$ can be related to the imaginary-time correlation function
\begin{equation}\label{current_current_correlation_function}
\Pi_{\alpha\beta}\left(\mathbf{q},\i\omega_n\right)=-\frac{1}{S}\int\limits_{0}^{\hbar\beta}\d\tau\left\langle\mathcal{T}\left[\hat{j}_\alpha(\mathbf{q},\tau)\hat{j}_\beta(-\mathbf{q},0)\right]\right\rangle\e^{\i\omega_n\tau}
\end{equation}
by $\Pi_{\alpha\beta}^R\left(\mathbf{q},\omega\right)=\Pi_{\alpha\beta}\left(\mathbf{q},\omega+\i0^\mathsmaller{+}\right)$, that is, by replacing $\i\omega_n$ with $\omega+\i0^\mathsmaller{+}$ in Eq.~(\ref{current_current_correlation_function}).\cite{Mahan2000,FetterWalecka2003,BruusFlensberg2006} Here, $\i\omega_n$ denotes a bosonic frequency. Hence, the Kubo formula for the real part of the conductivity reads as
\begin{equation}\label{real_conductivity}
\mathrm{Re}\left[\sigma_{\alpha\beta}\left(\mathbf{q},\omega\right)\right]=-\frac{\mathrm{Im}\left[\Pi^R_{\alpha\beta}(\mathbf{q},\omega)\right]}{\hbar\omega}.
\end{equation}

If vertex corrections due to phonons in Eq.~(\ref{current_current_correlation_function}) are ignored, the phonon-dressed Green's functions given by Eq.~(\ref{Greens_Function_via_self_energy}) expressed via their spectral functions~(\ref{spectral_function_general}), and the sum over $\mathbf{k}$ rewritten as a 2D integral, we arrive at
\begin{equation}\label{real_conductivity_final_q}
\begin{aligned}
\mathrm{Re}&\left[\sigma_{\alpha\beta}\left(\mathbf{q},\omega\right)\right]=\frac{2}{\hbar\omega}\sum\limits_{\lambda\lambda'}\int\frac{\d^2k}{(2\pi)^2}\int\frac{\d\omega'}{2\pi}\\
&\times D^{\mathsmaller{\lambda\lambda'}}_{\mathsmaller{\alpha\beta}}(\mathbf{k},\mathbf{q})\left[n_\mathsmaller{\mathrm{FD}}\left(\hbar\omega'\right)-n_\mathsmaller{\mathrm{FD}}\left(\hbar\omega+\hbar\omega'\right)\right]\\
&\times\mathcal{A}_{\lambda}\left(k,\omega'\right)\mathcal{A}_{\lambda'}\left(|\mathbf{k}+\mathbf{q}|,\omega'+\omega\right),
\end{aligned}
\end{equation}
where
\begin{equation}\label{dipole_product}
\begin{aligned}
D^{\mathsmaller{\lambda\lambda'}}_{\mathsmaller{\alpha\beta}}(\mathbf{k},\mathbf{q})&\equiv d^{\mathsmaller{K}}_{\mathsmaller{\lambda\lambda',\alpha}}(\mathbf{k},\mathbf{q})\left[d^{\mathsmaller{K}}_{\mathsmaller{\lambda\lambda',\beta}}(\mathbf{k},\mathbf{q})\right]^*\\
&=d^{\mathsmaller{K'}}_{\mathsmaller{\lambda\lambda',\alpha}}(\mathbf{k},\mathbf{q})\left[d^{\mathsmaller{K'}}_{\mathsmaller{\lambda\lambda',\beta}}(\mathbf{k},\mathbf{q})\right]^*
\end{aligned}
\end{equation}
is a real number.

Here, we are interested in the response to a uniform field, that is, in the case $\mathbf{q}=\mathbf{0}$, for which Eq.~(\ref{real_conductivity_final_q}) becomes
\begin{equation}\label{real_conductivity_final_zeroq}
\mathrm{Re}\left[\sigma_{\alpha\beta}\left(\mathbf{0},\omega\right)\right]=\delta_{\alpha\beta}\sigma\left(\omega\right)
\end{equation}
with $\sigma\left(\omega\right)$ given by Eq.~(\ref{real_conductivity_final_main}). In order to obtain Eqs.~(\ref{real_conductivity_final_main}) and~(\ref{real_conductivity_final_zeroq}), we have used that only $D^{\mathsmaller{\lambda\lambda'}}_{\mathsmaller{\alpha\beta}}(\mathbf{k},\mathbf{0})$ depends on the angle of the $k$-integration in Eq.~(\ref{real_conductivity_final_q}) for $\mathbf{q}=\mathbf{0}$ and that $\int\limits_{0}^{2\pi}\d\theta_\mathsmaller{{\mathbf{k}}}\,D^{\mathsmaller{\lambda\lambda'}}_{\mathsmaller{\alpha\beta}}(\mathbf{k},\mathbf{0})=\pi e^2v^2_\mathrm{\mathsmaller{F}}\delta_{\alpha\beta}$.

\bibliographystyle{apsrev}

\bibliography{BibOpticalConductivity}

\end{document}